\documentclass{article}
\usepackage[dvips]{epsfig}
\begin{document}
\rule{0cm}{1.5in}
\begin{center}
{\Large VCS above 8 GeV. The Experimental Challenge}
\footnote{Presented at {\bf Nuclear and Particle Physics
with CEBAF at Jefferson Lab, 3-10 November 1998, Dubrovnik,
Croatia.} }\\[0.5em]
{\large P.Y. BERTIN and Y. ROBLIN}\\
Universit\'e BLAISE PASCAL/IN2P3, Clermont-Ferrand FRANCE.\\
and\\
{\large C.E. HYDE-WRIGHT}\\
Old Dominion University, Norfolk VA USA.
\end{center}
\bigskip
\begin{abstract}
We discuss the experimental issues confronting measurements of 
the Virtual Compton Scattering (VCS) reaction
$e p \rightarrow e p \gamma$ 
with electron beam  energies 6--30 GeV.
We specifically address the kinematics of Deeply Virtual Compton
Scattering (Deep Inelastic  Scattering, with coincident
detection of the exclusive real photon nearly parallel to the
virtual photon direction) and large transverse momentum VCS
(High energy VCS of arbitrary $Q^2$, and the recoil proton emitted
with high momentum transverse to the virtual photon direction).
We discuss the experimental equipment necessary for these measurements.
For the DVCS, we emphasize the importance of the Bethe-Heitler --
Compton interference terms that can be measured with the
electron-positron (beam charge)
asymmetry, and the electron beam helicity asymmetry.
\end{abstract}

\section{ Introduction}
Exclusive reactions are a very powerful tool to study the transition between 
weakly interacting quarks at small distances and large distance effects such 
as the quark confinement responsible for the hadron structure.
One of the cleanest ways to tackle this problem is via  Virtual Compton 
Scattering (VCS), even though this reaction is
experimentally challenging for a number 
of reasons.
 First, the cross-section is small and  decreases as $Q^2$  or $s$
 increases. Furthermore, the VCS process ($e p \rightarrow e p \gamma$)
 has to be separated from a number of concurrent background processes,
 often with  counting rates several times higher than the VCS itself.
 Until recently this kind of experiment was not achievable because it 
requires  high beam current, high duty cycle, and low emittance.
The advent of CEBAF,
 enabled us to perform the first VCS experiment above the pion threshold at high Q$^2$ and s. This experiment is currently under analysis.\cite{ref:e93050}
 In the upcoming years this machine will be upgraded to 8 GeV, then 12 GeV and perhaps 24 GeV in the future. In the same time, the ELFE project in Europe is planning to use the existing LEP cavities to build a 30 GeV Machine with similar beam characteristics \cite{lep at CERN } (high current, high duty cycle and good energy resolution). 

\begin{figure}
\centerline{\epsfig{file=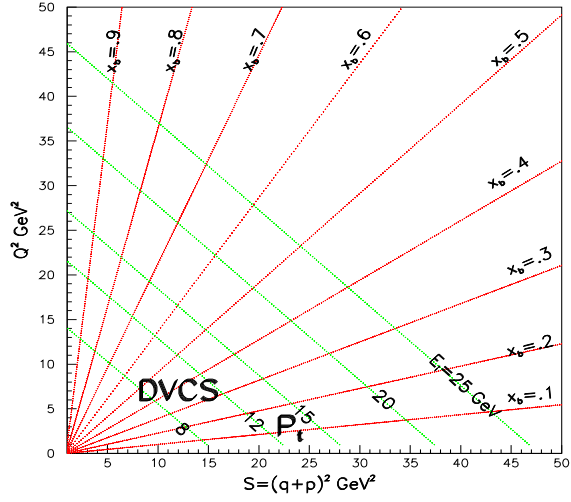,width=2truein,width=8 cm,height=7 cm}}
\caption{:
Ranges in Q$^2$ and $s$ accessible with an accelerator of incident energy E$_i$}
\label{Q2s}
\end{figure}

 In a contribution to the workshop CEBAF at 8 GeV \cite{8GeV et plus} we stressed the benefits of the VCS approach. This paper will focus on the experimental challenges for VCS experiments above 8 GeV.
  Our interest in VCS is twofold:
\begin{itemize}
  \item Deeply Virtual Compton Scattering (DVCS) corresponding to the 
  diffraction of a virtual photon in the forward direction. 
  DVCS\cite{x. Ji} \cite{A. Radyushkin} \cite{PAM Guichon and M. Vanderhagen} allows us to access the Off Forward Parton Distribution (OFPD) directly linked to the non perturbative part of the nucleon.
 The kinematic domain of DVCS is deep inelastic electron scattering
 (large $s$ and  Q$^2$) with the final photon produced very nearly in the virtual photon direction. For incident energies of 6 to 12 GeV, the accessible
domain is in the quark valence regime ($x_{\rm Bj} \approx 0.3$). 
  \item The Large P$_T$ domain. This is the domain where $s$ is large and the 
angle between the real and virtual photons is large (neither near 0 nor 
near $\pi$).   $Q^2$ is  moderate or even quasi-real.
 This probes the quark wave function  
\cite{PAM Guichon and M. Vanderhagen}\cite{P. Kroll}.
\end{itemize}
 Fig.~\ref{Q2s} shows the kinematical domains relevant for DVCS and Large $P_T$. It also shows the limit on $Q^2$ and $s$ given an accelerator energy.

\section{ Electro-production of a Photon}
\begin{figure}
\vskip3.5cm
\centerline{\epsfig{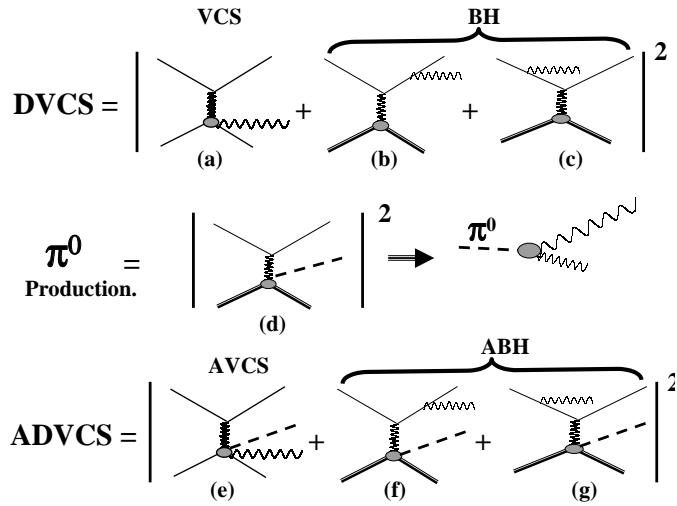}}
\vskip-0.5cm
\caption{Lowest order diagrams of electro production of photon, pion, or
photon plus pion.  Top: VCS and BH amplitudes. Middle: $\pi^0$ electro-
production, which contributes a background $e p \rightarrow e p \gamma \gamma$.
Bottom, Associated production:
 $e p \rightarrow e \gamma N^*$ $(N^*\rightarrow N\pi)$.
}
\label{diagrame}
\end{figure}

To the lowest  order in $\alpha$ the three graphs in the top panel of
Fig.~\ref{diagrame}  contribute to the
 electro-production of a  photon. 
Graph (a) is the VCS and
 graphs (b) and (c) are the Bethe-Heitler (BH) graphs. The latter two are fully calculable
 if we know the form factors of the proton, and their amplitude is purely real.

 Due to the propagator of the virtual electrons and the
  the virtual photon, three different poles have to be considered
in the  BH graphs: 
$$A^{\rm BH}\sim\frac{1}{t}\frac{1}{(k-q^\prime)^2-m_e^2}+
 \frac{1}{t}\frac{1}{(k^\prime+q^\prime)^2-m_e^2} $$
 These poles determine the shape of the BH cross section at high energy.
 Since the sign of the electron propagators are opposite in the two
 BH graphs there is a strong destructive interference at the $t=(q-q')^2$ pole.

\begin{figure}
\centerline{\epsfig{file=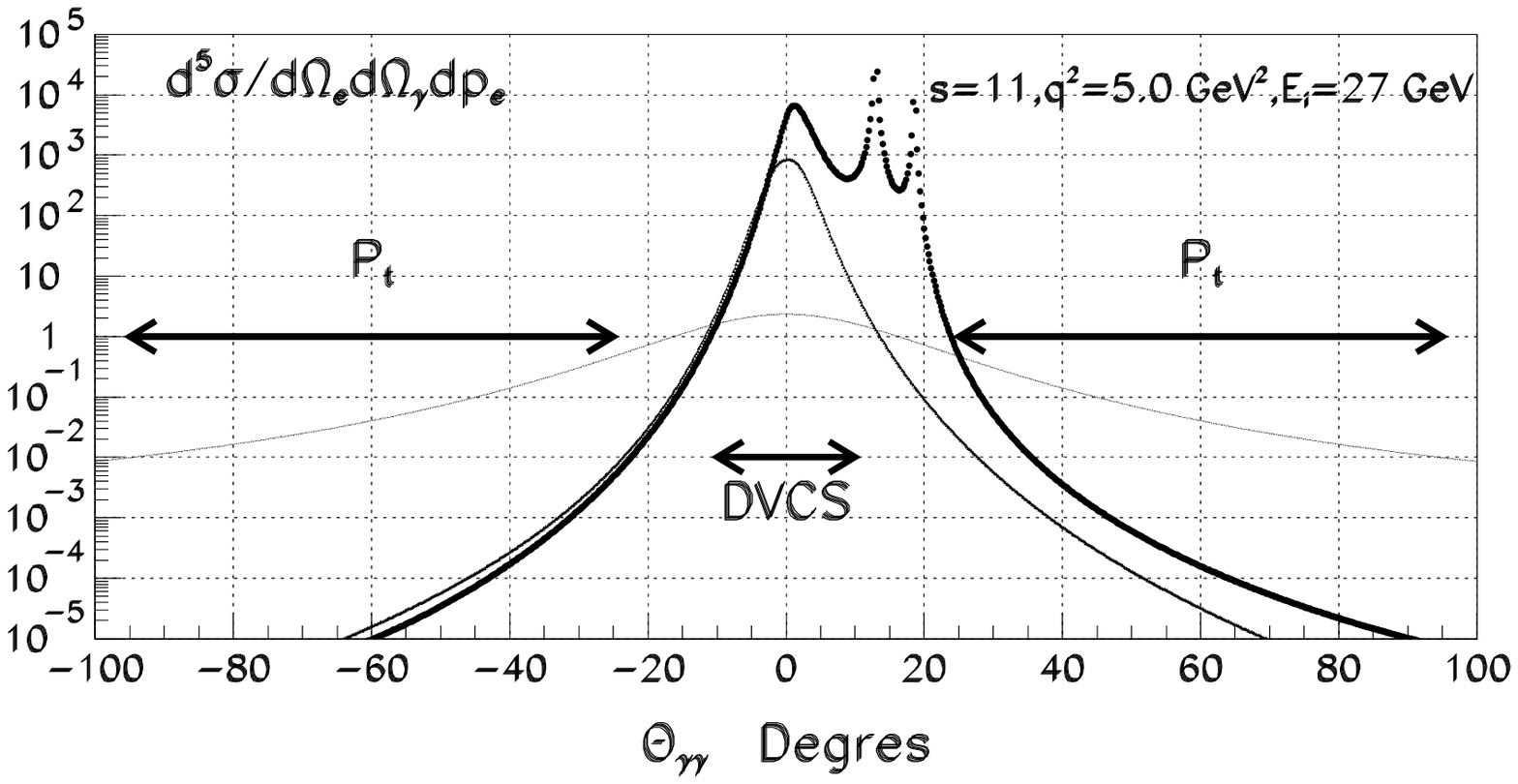,width=2truein,height=6 cm,width=12 cm}}
\vskip-0.3cm
\caption{Cross sections for the $e p \rightarrow e p \gamma$ process.
Heavy dotted line: Bethe-Heitler;  Medium line, DVCS; Thin line,
Model for Hard Scattering (Large $P_T$) VCS.  The cross sections are
plotted as a function of $\theta_{\gamma\gamma}$, the laboratory opening
angle between the virtual photon and emitted real photon directions.}
\label{vcsamp}
\end{figure}

\par

In Fig.\ref{vcsamp}, BH, DVCS, and large $P_T$ VCS cross sections
are evaluated for a particular electron kinematics.
The DVCS contribution is evaluated with a model given by P.A.M. Guichon and M. Vanderhaeghen \cite{PAM Guichon and M. Vanderhagen}. The OFPD are modeled
as the product of the DIS distribution functions and the elastic form factors.
The cross section for the large $P_T$ domain is modeled with the virtual
photon flux and a scaling ansatz for the photo-production cross section:
%
 $$ 2\pi\frac{d^5\sigma}{dp_e d\Omega_e dt d\phi_\gamma}=
{d\Gamma \over dp_e d\Omega_e} s^{-6}(30\ \mu\, b\cdot GeV^{10}).$$
 We will use these two models to estimate counting rates.
\par 

One of the main difficulties of VCS experiments is to be able to measure  
a cross-section over several (5$\sim$8) orders of magnitude. The dominant contribution
 is the BH at the two electron poles.  At the photon pole
 $\theta_{\gamma\gamma}=0$  ($t=t_{\rm min}$) the BH is  also  larger than VCS.
 Far away from these  poles BH becomes much smaller than the  VCS, and the cross section is small ($\sim$ 1 pb GeV$^{-1}$ sr$^{-2}$). 

The DVCS cross-section by itself is very small and the BH makes up most of the total cross-section in the DVCS regime. 
Fortunately, it is possible to extract the DVCS using asymmetries to access the interference term between the DVCS and the BH amplitude.  
This interference has an important contribution to the cross-section since the BH amplitude is large. This enhancement of the DVCS with the BH is the key to the measurement of the DVCS. There are two kinds of asymmetries that we can use:
\begin{itemize}
\item The lepton charge asymmetry, 
\item The beam polarization asymmetry
\end{itemize}

\subsection{Lepton charge  asymmetry}

  This asymmetry is measured by the difference in the cross section 
for a negative or positive incident lepton (electron to positron):
\begin{itemize}
\item The VCS amplitude (diagram (a), Fig.~\ref{diagrame})
  $T_{VCS}$ is anti-symmetrical under a charge conjugation, there is  only one coupling on the lepton line.
\item
The BH amplitude (diagrams (b and c), Fig.~\ref{diagrame})  $T_{BH}$  is symmetrical, there are two couplings onto the lepton line.
\end{itemize}
 Therefore, the interference term between the BH and the VCS in the cross section is the only term contributing to an  electron-positron asymmetry \cite{PAM Guichon and M. Vanderhagen} 
    $$d^5\sigma^{e^-}-d^5\sigma^{e^+}=\sim  4\Re [T_{VCS}\cdot T_{BH}]$$
 This asymmetry is a direct measure of the VCS amplitude since the BH amplitude is
fully calculable.

\begin{figure}
\centerline{\epsfig{file=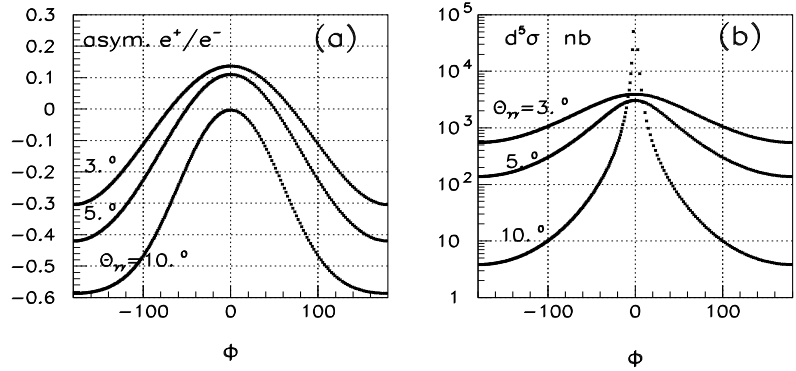,width=2truein,height=6 cm,width=12 cm}}
\vskip-0.5cm
\caption{{\bf(a)} Asymmetry versus azimuthal  angle  $\phi$
induced when the sign of the beam is changed
  (electron/positron beam). The plot is given at three angles
 $\theta_{\gamma\gamma} = 3^\circ$, $5^\circ$ and $10^\circ$.
 {\bf (b)} The cross section 
$d\sigma /  (dk_e d\Omega_e d\Omega^{\rm lab}_{\gamma\gamma})$
 in nb / (GeV sr$^2$) at the same angle.
For $\phi = 0$, the emitted photon lies in the scattering plane, 
closer to the beam than the virtual photon direction.
The curves are calculated with the model of 
Ref.~\cite{PAM Guichon and M. Vanderhagen}.
 We have chosen the kinematic at $s = 8$ GeV$^2$,
 Q$^2$=3 GeV$^2$and an incident Energy $E_i = 8$ GeV. }
\label{Asymmetry ee}
\end{figure}

We use in the following the asymmetry:
 $$A^{e^+/e^-}= \frac{d^5\sigma^{e^-}-d^5\sigma^{e^+}}{d^5\sigma^{e^-}+d^5\sigma^{e^+}}.$$
We give in Fig.~\ref{Asymmetry ee} the value of this asymmetry for three 
angles between the virtual photon and the real photon $\theta_{\gamma\gamma}$.
 It is plotted against the azimuthal angle $\phi$ between the leptonic and the 
 hadronic plane. This figure is for an  incident beam energy of 8 GeV, s=8 GeV$^2$ and $Q^2$= 3 GeV$^2$.
From the figure we see that   the asymmetry is large for
small angles of the real photon relative to the virtual photon direction.
 
 We think it is very interesting to use this characteristic and that is 
 why we propose to build
 at CEBAF and at ELFE a positron beam.  We will come back on this point later.

\subsection{The beam polarization asymmetry.}

  Another interesting observable is the beam helicity asymmetry 
$$A^{Beam}= \frac{d^5\sigma^{\rightarrow}-d^5\sigma^{\leftarrow}}{d^5\sigma^{\rightarrow}+d^5\sigma^{\leftarrow}}$$
 produced with a polarized beam.  This asymmetry is defined by the ratio between the difference and the sum of the cross-sections obtained when reversing the beam longitudinal polarization.
 Here also, the interference between the BH amplitude and the VCS is the 
only term contributing (assuming the longitudinal VCS amplitude is
much smaller than the BH).

 This was pointed out in the case of large $P_T$ by P. Kroll et al.  
\cite{P. Kroll}\cite{PAM Guichon and M. Vanderhagen}{}
 and afterwards applied by M. Diehl et al  \cite{M.Diehl Ralston and B. Pire} and PAM Guichon and M. Vanderhaeghen \cite{PAM Guichon and M. Vanderhagen} in the case of
 the DVCS. 

\par

\begin{figure}
\centerline{\epsfig{file=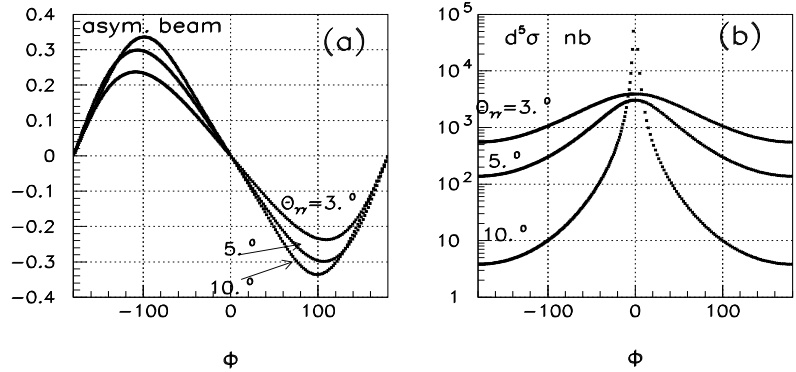,width=2truein,height=6 cm,width=12 cm}}
\vskip-0.5cm
\caption{{\bf(a)} Asymmetry versus angle induced when the sign of the longitudinal polarization of the
  beam is reversed . The plot is given at three angles $\theta_{\gamma\gamma}$ 3, 5 and 10 degrees. {\bf (b)} 
The cross section at the same angle. We have chosen the kinematic at s=8 GeV$^2$, Q$^2$=3 GeV$^2$ and an incident
 energy E$_i$=8 GeV. }
\label{Asymmetry phi}
\end{figure} 

 We give in Fig.~\ref{Asymmetry phi} the value of this asymmetry, for the same  kinematics as in Fig.~\ref{Asymmetry ee}.
The asymmetry deviates slightly from a pure $\sin\phi$ behavior, due to
the structure of the BH amplitude. This asymmetry is maximal out of 
plane ($\phi=\pm 90\deg $) and
 is zero  when the photon angle $\theta_{\gamma\gamma}$ is zero.
Unfortunately the cross section is maximum when  $\theta_{\gamma\gamma}$ is
zero. 
The beam helicity  asymmetry will require a larger integrated luminosity
to achieve the same precision as the beam charge asymmetry.

\section{Experimental Equipment}

In order to select VCS events (DVCS or large P$_T$) 
 one must make sure to select photon electro-production events.
 To do that we can use the  squared  missing mass.  We need to know:
\begin{itemize}
\item the incident particle  - that is why it is so crucial to have a good quality beam;
\item  the scattered electron - it fixes the virtual photon;
\item   the recoil proton and/or the photon produced. The choice to detect the recoil proton and/or
  the photon will be fixed by the kinematics and the level of resolution needed.
\end{itemize}
 If the photon is not detected, it is necessary to separate the missing  mass 
 zero  (photon missing mass) from the pion  mass. The  squared missing mass  resolution must be $$ \Delta M_X^2\leq M_\pi^2.$$
 If we measure the photon (instead of the proton),
 the resolution we require on  $M_X^2$ is much looser:
 $$ \Delta M_X^2\leq (M_P+M_\pi^2)-M_p^2\sim 2 M_pM_\pi.$$ There is a 
factor of twenty between the required resolution on the squared missing mass in the two cases.

The best case will be when  the photon is detected and measured  (momentum and direction) and the proton detected  (position only). In this case we can use
not  only a missing mass technique, but also require  coplanarity conditions on the virtual photon, the recoil proton and the photon.

\subsection{The Background Problem}
 There are two main sources of background. The $\pi^0$ electroproduction:
$e p \rightarrow e p \pi^0$ (diagram d, figure \ref{diagrame}),
 and the pion associated production with photon electroproduction:
$ e p \rightarrow e \gamma N^*$ (diagrams e,f and g figure \ref{diagrame}).

\subsubsection{The $\pi^0$ electro-production.}

 The graph of the  $\pi^0$ electro-production is the d graph of the
  figure \ref{diagrame}.
 When the $\pi^0$ decays with a photon emitted
 in the forward direction, the second photon from the $\pi^0$ is backward and has
 a very low energy (few MeV). The final products of this reaction are nearly
 the same as a VCS event except for a soft backward photon which
is very difficult to detect 
 at an electromagnetic machine. The missing mass 
technique is
unable to solve the problem either. 
The $\pi^0$ events are partially removed by a coplanarity cut  
on triple $e p \gamma$ coincident events. 
The solution for the remaining $\pi^0$ events
is to record in the calorimeter events with the
  $\pi^0$ decay at 90 degrees in the centre of mass. The two photons
 are emitted in the forward direction with  comparable energies and
their opening
angle is  $\theta^{\pi}_{\gamma\gamma} \ge\frac{2\,m_{\pi^0}}{P_{\pi^0}}$.
Using these events we can infer the $\pi^0$ cross section and subtract its 
 contribution from the events with only one photon recorded.

\begin{figure}[ht]
\centerline{\epsfig{file=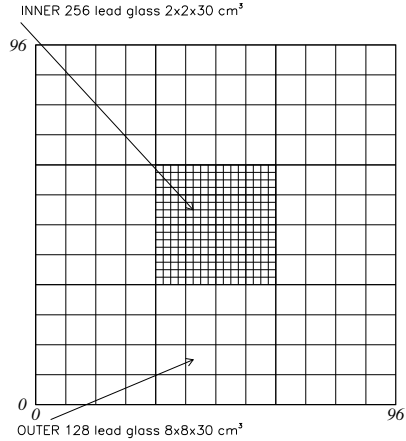,height=6 cm,width=6 cm}}
\caption{Calorimeter concept with high inner granularity to record
DVCS events and coarser outer granularity to record two photon
events from the $e p \rightarrow e p \pi^0$ reaction.}
\label{inner-outer}
\end{figure}
 
\par
In the figure \ref{inner-outer} we give 
an example of a composite calorimeter. The inner part, with a small
 granularity to collect the VCS events,  the outer part with a bigger granularity
 to collect the two photons events from the $\pi^0$ decay.
\par

  In the DVCS kinematics the $\pi^0$  production:
\begin{itemize}
\item  Decreases as 
$1/Q^6$, faster than the DVCS which decreases as $1/Q^4$.
\item Has no interference with BH, 
contrary to the VCS
 which is amplified by the BH when the quadri-transfer $t$ is small. 
\item Has no beam charge-dependent asymmetry. Thus this background
vanishes identically in $A^{e^+/e^-}$.
\end{itemize}
\par
 We should point out that the $\pi^0$ cross section is in itself a  very interesting result.
 In DVCS kinematics, it is sensitive to another 
combination of the off forward parton distribution \cite{Guidal}. 
\par
 
\subsubsection{The Associated Pion Production.}
  Another parasitic reaction is the associated pion production 
 at  the photon electro-production. The corresponding graphs are 
 Fig.~\ref{diagrame} e, f and g. The
 last two graphs (f and g) 
 are the associated pion production with the BH  
 process (ABH).
The A-BH process can be exactly predicted,
given knowledge of the $p\rightarrow N^*$ transition form factors
(instead of the proton form factors as in the VCS).
The third graph (AVCS) is the pion production in the VCS. 
It leads to the same final state as the  e and f graphs
 and therefore the three graphs interfere. 

 The physics of associated production in DVCS
(f) is just as important as the physics of the DVCS process.
For example, for  pion-nucleon system at the mass of the
 $\Delta$ (or higher $N^*$) A-DVCS gives access to some of the
$N\rightarrow N^*$ transition OFPD's.
 However it is not so simple to extract the A-DVCS amplitude since,
 in this case A-BH is no longer purely real but has an imaginary part as well
 (because of the intermediate on-shell $N^*$ state).
 Close to the pion threshold  ($\pi N$ system in s-wave)
 a low energy theorem can be
 built to relate the $p\rightarrow N\pi$ OFPDs to the
elastic OFPDs.
\par
 Under the charge change of the lepton, 
the associated pion production has an asymmetry similar to the DVCS case. 
\par

 If the proton is not detected, the associated  pion production can
 appear in either the $\pi^0P$ or $\pi^+ n$ final states. If the
recoil proton is detected,
 only the $\pi^0$  channel is open.

\begin{figure}
\centerline{\epsfig{file=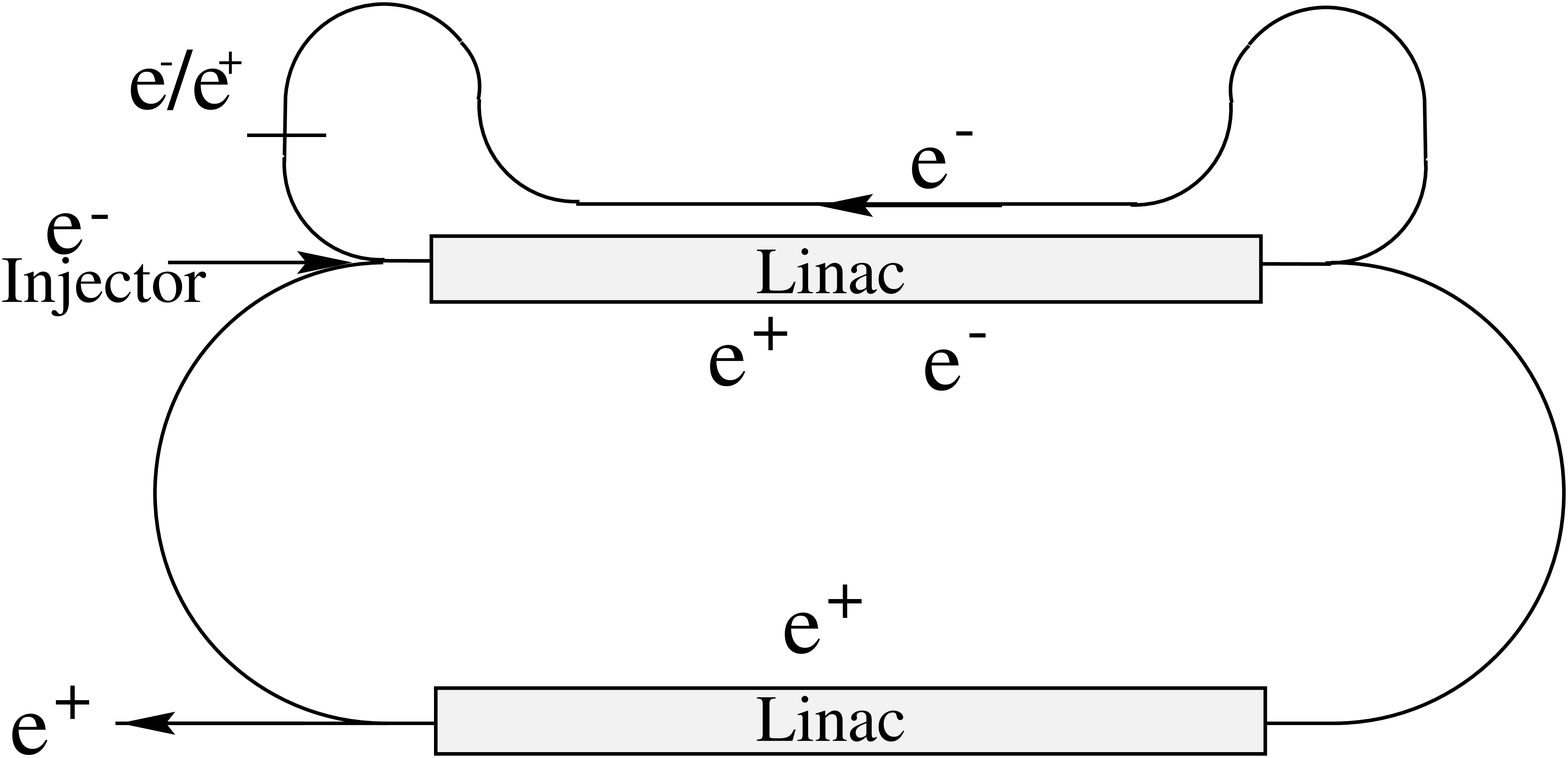,width=2truein,height=4 cm,width=10 cm}}
\caption{Possible setup of a positron beam at CEBAF.}
\label{positron}
\end{figure}

\subsection{Positron beam}
\label{pbeam}
 We have pointed out the interest of the charge asymmetry. But to access
 this information we need a positron beam. 
Several techniques can be used such as  radioactive sources but the main
 technique is to produce positrons on a heavy material target. This yields
 positrons of about 60 MeV, which are then collected and injected into the Linac. This does not require any significant changes
 in the Linac setup. (Fig.~\ref{positron}). 
 Positrons can just be injected into
 the Linac with a RF phase difference of 180$^\circ$ 
 relative to the electrons. The positron yield is proportional 
 to the electron beam power on the conversion
 target. Table~\ref{positron beam}\cite{ H. Tang}{}
  lists the parameters
 of the positron source used at the SLC, the parameters
 of the source for the Next  Linear Collider (NLC).
   The CEBAF extrapolation is  obtained by scaling the
 to the CEBAF beam power on the conversion target.
%
\cite{ H. Tang}.
 We have given for CEBAF 
two extrapolations  (a) and (b) based  the SLC number and the NLC Project. 
This latter give a larger positron yield due to an improved
 positron
 collection.
  We can conclude that a positron beam is realistic for CEBAF. 
Because the asymmetry is large we do not need rapid switching from
electrons to positrons.  We need to switch the beam (reverse the
polarity of the arc magnets) approximately once every 100 hours.

 The following modifications needs to be done on the beam line to provide a positron beam:

\begin{itemize}
\item A new magnet at the end of the north Linac,
\item a beam transport line in the north tunnel,
\item a room for the positron target, and the optics to collect the positrons,
\item a 80 KW beam dump for 0.5 GeV beam
\end{itemize}
   
\begin{center}
\begin{table}
\begin{tabular}{||l|c|c|c||} \hline \hline
{\bf Parameters}&SLC94&NLC-II&CEBAF \\ \hline \hline
Electron Beam Drive&&& \\ \hline 
\ \ \ \ Electron Energy (GeV)&30&6.22&0.5 \\ \hline 
\ \ \ \ Bunches by pulse &1&75& \\ \hline 
\ \ \ \ Repetition Rate (Hz)&120&120& \\ \hline 
\ \ \ \ Bunch Intensity ($e^-$) & $3.5\cdot 10^{10}$&
                                       $1.5\cdot 10^{10}$& \\ \hline 
\ \ \ \ Pulse Intensity ($e^-$) & $3.5\cdot 10^{10}$&
                                       $113.\cdot 10^{10}$& \\ \hline 
\ \ \ \ Intensity  ($e^-/$ sec)      & $4.20\cdot 10^{12}$
                    &$1.35\cdot 10^{14}$&$1.00\cdot 10^{15}$ \\ \hline 
\ \ \ \ Intensity ($\mu$A)&0.67&21.6&160 \\ \hline 
\ \ \ \ Beam Power (KW)&20.2&134&80 \\ \hline 
Positron Collection&&& \\ \hline 
\ \ \ \ Yield (e$^+$ per e$^-$)&2.4&2.1&$\sim$ (a) 0.04 (b) 0.17  \\ \hline 
\ \ \ \ Intensity ($10^{14} e^-/$s)&0.108 &2.84 &
(a) 0.4  (b) 4.3  \\ \hline 
\ \ \ \ Intensity ($\mu$A)&1.7&45.8&$\sim$ (a) 6.4 (b) 27 \\ \hline \hline 
\end{tabular} 
\caption {Positron production 
from the SLC-94 and  the Next Linac Collider II\cite{H. Tang}.
The estimation for
 CEBAF is obtained by  scaling   with the CEBAF beam power 
 (a)  on SLC, or  (b) on the future project NLC-II, which increases
 the parameter of 
the positron collection.   }
\label{positron beam}
\end{table}
\end{center}

\subsection{Electron spectrometer}
 
When one detects the scattered electron one must overcome two difficulties:
\begin{itemize}
\item Reach small angles to do the physics of interest
\item Have enough solid angle to insure high counting rate.
\end{itemize}
This must be done keeping a high momentum acceptance
 ($\frac{\Delta P}{P}=5-10 \%$) and a momentum resolution of  10$^{-4}$.

It is not difficult to go at small angles but this is often at the expense of the solid angle.
For example,  experiment E154 at SLAC reach $\theta_e$ = 5.5$\deg$ with a solid angle $\Omega$= .5 msr, and   $\theta_e$=2.75$\deg$ with  $\Omega$= .15 msr. Several  solutions are possible to boost the solid angle while still being at small angles: 
\begin{itemize}
\item use of $\cos n\theta$ magnets such as was proposed by P, Vernin et al. for
 the ELFE project \cite{P.Vernin ELFE} or  in this conference \cite{P. Vernin Dubrovnique}or by J.M. Finn
 \& Al. \cite{M. Finn 8 GeV+}  for 8-12 GeV at CEBAF.
 This solution gives a good
 spectrometer acceptance  ($\Omega$=6.8 msr) and $\frac{\Delta P}{P}=20 \% $ 
in the case of ELFE project at 15 GeV . 
\item use of Septum magnet. This is currently being build by the INFN group for the CEBAF HRS. It
will be possible to reach 6$\deg$ with a solid angle of $\Omega$=6 msr.
\item If a smaller angle is desired  (smaller then 1$\deg$), this can be done using a setup using quadrupoles, like the 
M\o ller setup in the experimental hall A.
\end{itemize}

\subsection{Photon calorimeter}

The  specifications  of the photon calorimeter are very different if we consider physics at 
large momentum transfer or DVCS kinematics. The next two sections list the requirements on the apparatus for each type of experiment.

\subsubsection{ Large $P_T$ $H(e,e'p\gamma)$ }
In this case the main purpose of the calorimeter is to suppress
 the accidentals in the reaction. It must be used at very high
luminosity ($1-2 \cdot10^{+38}$ cm$^{-2}$s$^{-1}$) and will be placed at 
large angle (50 $\deg$). It does not need high energy resolution 
(30-50 \%), rather it must have a large acceptance coverage (1.5 sr). 
This  calorimeter can be shielded from the low energy gamma rays, and can be
 close to the target.  It can be built  in lead/chamber or lead/plastic sandwich. 
The fact that it must be able to remove accidentals requires a moderate granularity. 
That granularity needs to be somewhat better if it is necessary to identify the $\pi^0$ decay.
The two decay photons are separated by an angle of 
$\theta_{\gamma\gamma}=\frac{2m_{\pi^0}}{P_{\pi^0}}$. 
This  is roughly two degrees 
 at $s=20$ GeV$^2$, $Q^2 = 5$ GeV$^2$ and an incident energy of 27 GeV, the 
photon energy at the maximum $P_T$=2.1 GeV is $q'=6.6$ GeV. If the granularity is good enough one can also 
use it for coplanarity cuts (requiring that the detected photon, the recoil proton and the VCS virtual photon all lie
 in the same plane)

 \subsubsection{DVCS}

  In this case the calorimeter will be placed in a forward direction
  (10--20 $\deg$)  where the electro magnetic background coming from 
the target is a limiting factor for the luminosity, 
a few $10^{37}$ cm$^{-2}$s$^{-1}$ 
instead of $10^{38}$ cm$^{-2}$s$^{-1}$.
  The calorimeter solid angle is also much smaller (10 msr) 
  than in the case of large $P_T$. This is just
 a fact of the Jacobian.  But if we want solve the missing photon from
 the $\pi^0$ decay  this acceptance must be increased  to cover the two photons decay of the  $\pi^0$  :
$$\Omega_{\gamma\gamma}^{\pi}\gg (\frac{2m_{\pi^0}}{P_{\pi^0}})^2.$$

\begin{table}
\begin{center}
\begin{tabular}{||r|r|r|l|l|r||} \hline \hline 
s      & Q$^2$ & E$_i$ &$P_{\gamma}$ & $\Omega$ &$\Delta \Theta$ \\
GeV$^2$&GeV$^2$&GeV    &GeV  &  msr     & mr            \\ \hline
 6     &   1   &  6    & 3.2 & 7.4      & 86.          \\ \hline
 8     &   3   &  8    & 5.3 & 2.7      & 52.          \\ \hline
 11    &   5   &  16   & 8.0 & 1.3      & 36.          \\ \hline
 15    &   7   &  27   &11.2 & 0.6      & 25.         \\ \hline \hline
\end{tabular}
\end{center}
\caption{Minimum Calorimeter  acceptance $\Omega$ required to measure the $\pi^0$ electro-production
 cross section to subtract it from the DVCS. We also give the required 
 separating power $\Delta \Theta$ of two photon in the calorimeter to resolve both photons in $\pi^0 \rightarrow \, \gamma \gamma$.}
\end{table}     
 
One must stress again that if we work with charge asymmetry,  
(elec\-tron-positron)
 then the electro-production of  $\pi^0$ is gone (it does not contribute to the asymmetry)  and our only worry is the associated pion production.
 To distinguish between  the associated pion production and the  $H(e,e'\gamma)p$ reaction we need an energy resolution on the photon:
$$\Delta P_{\gamma}\ll \frac{M_pm_\pi}{P_X}$$
or we have to detect the recoil proton to check for coplanarity.

\begin{figure}
\centerline{\epsfig{file=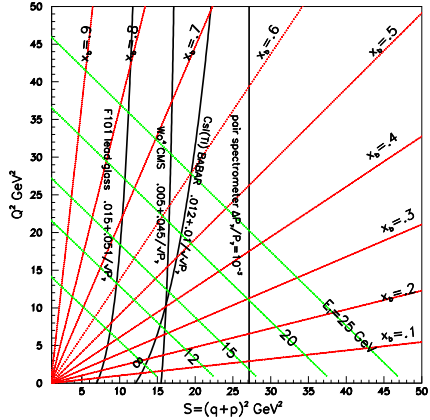,width=2truein,height=6 cm,width=6 cm}}
\vskip -0.3cm
\caption{Photon resolution on a plot $Q^2$ vs. $s$.}
\label{calo_reso}
\end{figure} 

In the figure \ref{calo_reso}
we give on a plot ($s$,$Q^2$) 
the required photon resolution for one sigma separation of the $p$ and $N\pi$
  final states in the 
$H(e,e'\gamma)X$ reaction.  
In order to estimate the precision of  the energy measurement
of the photon, we consider three kind of crystals:
\begin{itemize}
\item Lead tungstate  used on the CMS  calorimeter \cite{lead} giving a resolution when the crystal is seen by APD:
$$\frac{\sigma_E}{E}=.035+\frac{.036}{\sqrt{E}}$$
\item F101 Radiation Resistant Lead Glass \cite{F10}. The resolution 
$$\frac{\sigma_E}{E}=.015+\frac{.051}{\sqrt{E}}$$
is obtained with a preshower compensation.
\item 
CsI(Ti) used by BaBar at SLAC \cite{CSTI}. The resolution is 
$$\frac{\sigma_E}{E}=.012+\frac{.01}{\sqrt[4]{E}}$$
 This calorimeter is without any doubt one of the best on the market. 
If we want to go at higher energy we can use a pair spectrometer
 with a converter and a magnetic field.
\end{itemize}
  From the figure \ref{calo_reso} we can see that at moderate $s$
 (smaller then 10 GeV$^2$) it is
 possible to build a calorimeter meeting all the specifications for the DVCS.
 
\par
    High luminosity will  be the main issue for the calorimeter, mainly in the forward angle
 where the electro-magnetic background coming from the target will  be large 
(M\o ller, Radiative M\o ller, 
scattered electrons...). Several tests done in  HALL A for the RCS experiment
97-108 have
 already shown that working at a luminosity of few 10$^{37}$ cm$^{-2}$ s$^{-1}$ is possible.
 In order to deal with the pile-up in the calorimeter we will use a new technology based on high speed sampling (1 GHz) on the calorimeter channels.
 
\subsection{Proton detector} 

  There are two different approaches for the proton detectors depending on what we want to measure: 
\begin{itemize}
\item
 In the case of VCS at Large $P_T$ we use the reconstructed proton momentum
 to build the missing mass and identify the reaction. This means one needs a 
high resolution spectrometer (the same level of performances one has in the electron arm). A design with  $\cos n\theta$ magnets  is discussed
 in \cite{P.Vernin ELFE},\cite{P. Vernin Dubrovnique}.
\item 
 In the case of DVCS we may only need to perform coplanarity tests. 
It is possible to chose kinematics such that the angle $\theta^{lab}_{P,\gamma^v}$ between the recoil proton and the virtual photon in the laboratory frame is large.
 The Center of mass angle corresponding to this kinematic is small and this is a DVCS  kinematics   (small t).  
The figure \ref{angle proton_photon_lab} shows
 some  possible kinematics at several incident energies and  Q$^2$ and S.

\begin{figure}
\centerline{\epsfig{file=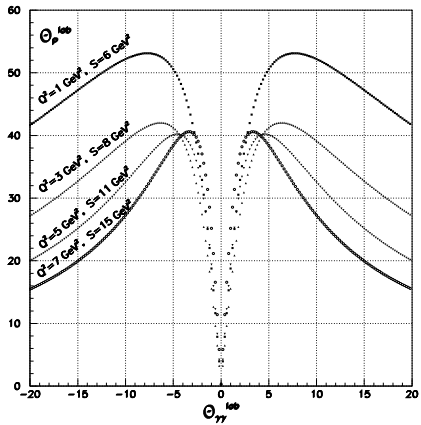,width=2truein,height=6 cm,width=6 cm}}
\caption{Recoil proton angle  versus
 emitted photon angle in the laboratory frame for the H$(e,e'p\gamma)$
 reaction.
Both angles are measured relative to the virtual photon direction.}
\label{angle proton_photon_lab}
\end{figure} 

\begin{table}
\begin{center}
\begin{tabular}{||r|r|r|r|r|l|r||} \hline \hline 
s      & Q$^2$ & E$_i$ &$\theta_{\rm cm}$& $\theta_{p\gamma}$&$P_P$  & $\theta_{\gamma\gamma^v}$ \\
GeV$^2$&GeV$^2$&GeV    &deg.  &deg     & GeV&deg.            \\ \hline
 6     &   1   &  6    & 23. & 53. & 0.51    & 7.5          \\ \hline
 6     &   2   &  6    & 30. & 42. & 0.77    & 8.5          \\ \hline
 8     &   3   &  8    & 27. & 42. & 0.85    & 6.5          \\ \hline
 11    &   5   &  16   & 23. & 40. & 0.94    & 4.6          \\ \hline
 15    &   7   &  27   & 20. & 40. & 0.98    & 3.4         \\ \hline \hline
\end{tabular}
\end{center}
\caption{ Kinematics of the maximum angle  $\theta^{Lab}_{P\gamma^v}$ between the proton and the virtual photon, in the H$(e,e'p\gamma)$
 reaction.}
\label{anglemaxkine}
\end{table}

Table~\ref{anglemaxkine} gives several examples
 of DVCS kinematic  computed at the maximum angle in the lab between the proton
 and the virtual photon. At this angle the Jacobian
 $J=\frac{d\, \Omega^{cm}_P}{d\, \Omega^{lab}_P}$ is maximal 
(meaning our CM solid angle is maximal).
 
From this table several conclusions can be extracted:

\begin{enumerate}
\item The proton momentum is high. In this table, the lowest  P$_P$=.551 GeV has
 a range of 20 g cm$^{-2}$ in iron.   
 This implies that we can shield the proton detector from low energy 
 particles without losing in efficiency on the proton detection. Since most 
 of the high noise counting rate is at low energy this also has the benefit to allow us to use high luminosities.

\item The angle  between the virtual photon and  the real photon  is small.
 Thus, we can place
 the photon calorimeter in the direction of the virtual photon. This 
calorimeter can have a small angular acceptance and we will still be able to catch the real photon.
 The optimum position is a tradeoff between this angle and the electron spectrometer acceptance. 
\item The angle of the proton and the virtual photon is large (40 degrees for the smallest
 in the table). The virtual photon is close to the forward direction, 
 (10-20 degrees) the proton detector however will be farther thus, less sensitive to background 
 allowing us to work at high luminosities.
\end{enumerate}
 \item The solution proposed for the photon calorimeter to solve the pile-up problem based on sampling at 1 Gigahertz will nicely improve the detection of the proton also.
\end{itemize}

\begin{figure}
\centerline{\epsfig{file=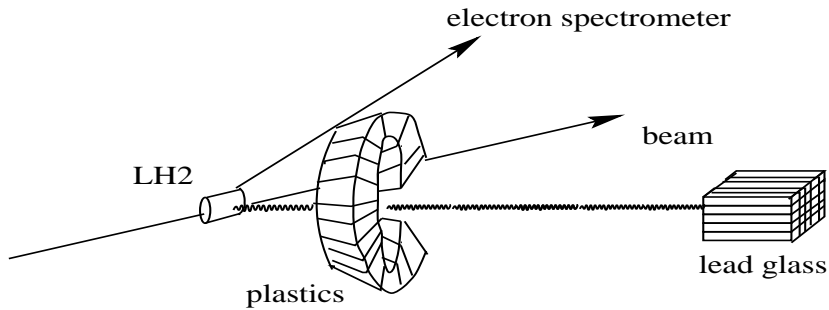,width=2truein,width=6 cm,height=12 cm,angle=-90}}
\vskip-1.0cm
\caption{Proposed experimental setup for DVCS.}
\label{setup}
\end{figure} 

 Finally,  we can propose the  experimental setup sketch presented in the figure \ref{setup}.
The proton detector is a ring of plastic scintillators located around
 the direction of the virtual photon. The ring is partially open in the forward direction to let 
the beam and the scattered electron  through.

\begin{figure}
\vskip-0.2cm
\centerline{\epsfig{file=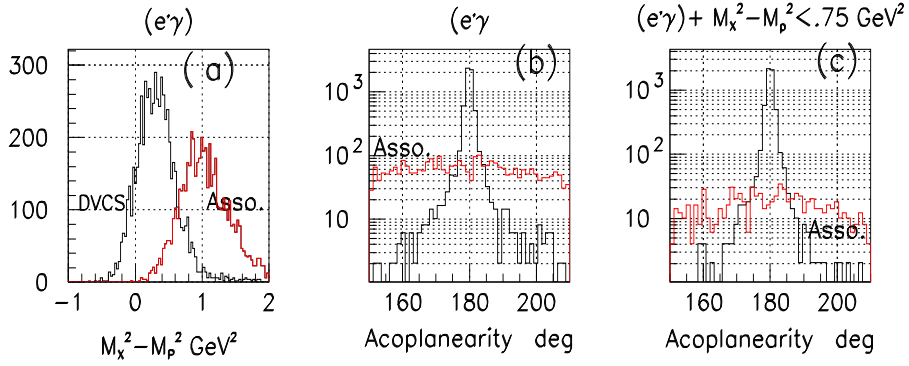,width=2truein,width=13 cm,height=6.5 cm}}
\vskip-0.7cm
\caption{ Missing mass plot  and acoplanarity spectra for the
$e + p \rightarrow e + \gamma + p $ and 
$e + p \rightarrow e +  \gamma + N\pi$ reactions. 
Kinematic is at Q$^2$=3 GeV$^2$, S=8 GeV$^2$,  8 GeV incident energy. The electron spectrometer HRS is at 19.9 Degrees with a solid angle of 6 msr. The inner photon calorimeter is located at -11.9 degrees and its solid angle is 40 msr. The density of the events on this plot is just given by the phase space, there is no cross-section weighting.
{\bf a)} Missing mass squared for H$(e,e'\gamma)X$.  {\bf b)} Out of plane
angle between final photon and proton for H$(e,e'\gamma p)$
and H$(e,e'\gamma p)\pi^0$ reactions. {\bf c)} Same as b), but with
cut on missing mass spectrum of a). }
\label{cut_ei8}
\end{figure} 

\begin{figure}
\vskip-0.4cm
\centerline{\epsfig{file=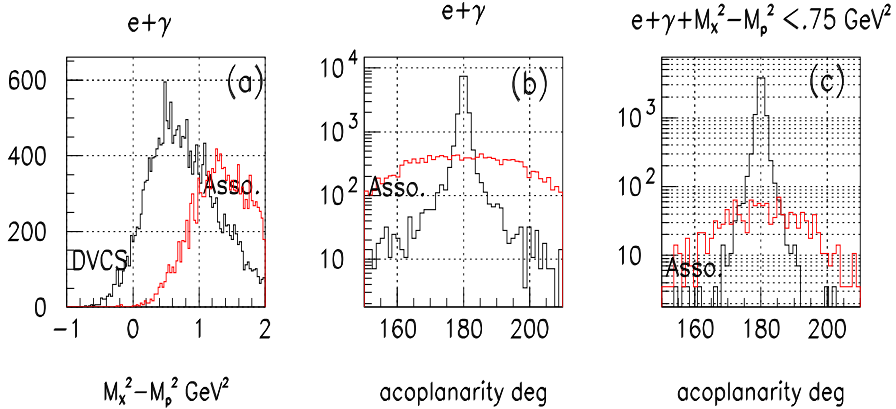,width=2truein,width=13 cm,height=6.5 cm}}
\vskip-0.5cm
\caption{ Missing mass plot and acoplanarity spectrum. 
Kinematic is at Q$^2$=5 GeV$^2$, S=11 GeV$^2$,  27 GeV energy incident. The electron spectrometer is  a new spectrometer at 5.7 degrees with a solid angle of 2 msr. The inner photon calorimeter is located at $-10.0$ degrees and this solid angle is 40 msr.
The density of the events on this plot is just given by the phase space, there is no cross-section weighting. The plots are the same as in 
Fig.~\ref{cut_ei8}. }
\label{cut27}
\end{figure}

In Figs.~\ref{cut_ei8},\ref{cut27}{} 
the plots a) shows the separation between $p(e,e'\gamma )p$ and
$p(e,e'\gamma)N^*$ achievable  using only  the electron and the 
photon information. The plots b) \& c) show the discriminatory power
of the coplanarity distribution using a proton array
to measure  $p(e,e'\gamma p)X$ events.  The plots b) \& c) are 
without and with a cut on the missing mass obtained in a).



\begin{figure}
\centerline{\epsfig{file=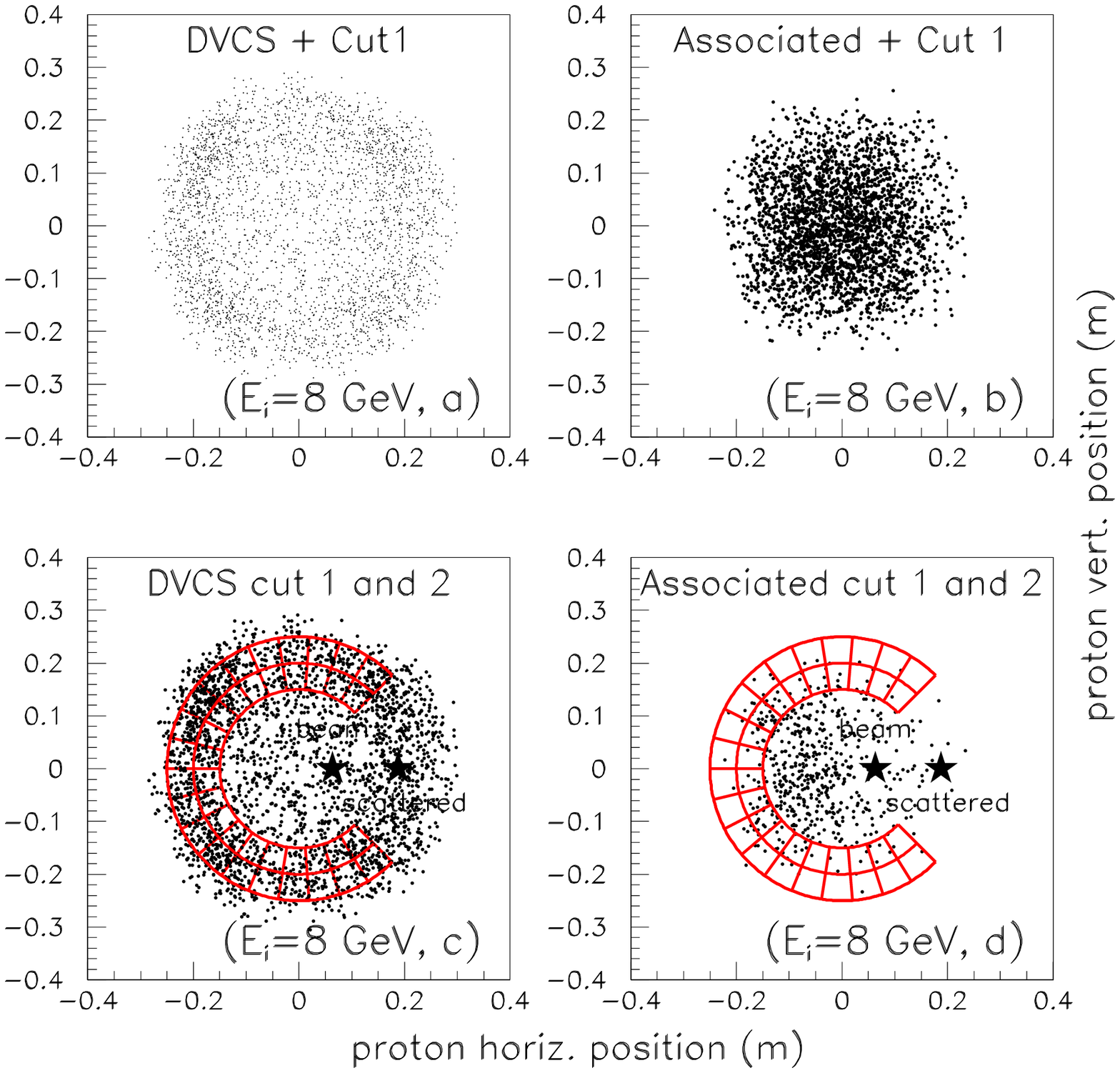,width=12 cm,height=12 cm}}
\caption{ Localization of the proton in a plane perpendicular to
the mean virtual photon direction, 
30 cm from the LH2 Target.
Kinematic is at Q$^2$=3GeV$^2$, S=8 GeV$^2$,  8 GeV energy incident. The electron spectrometer HMS is at 19.9 Degree with a solid angle of 8 msr. The inner photon calorimeter is located at -11.9 degrees and this solid angle is 40 msr. Note that the density of the event is just given by the phase space.  The proposed array of proton detectors is also illustrated. }
\label{calo_position_8}
\end{figure} 

\begin{figure}
\centerline{\epsfig{file=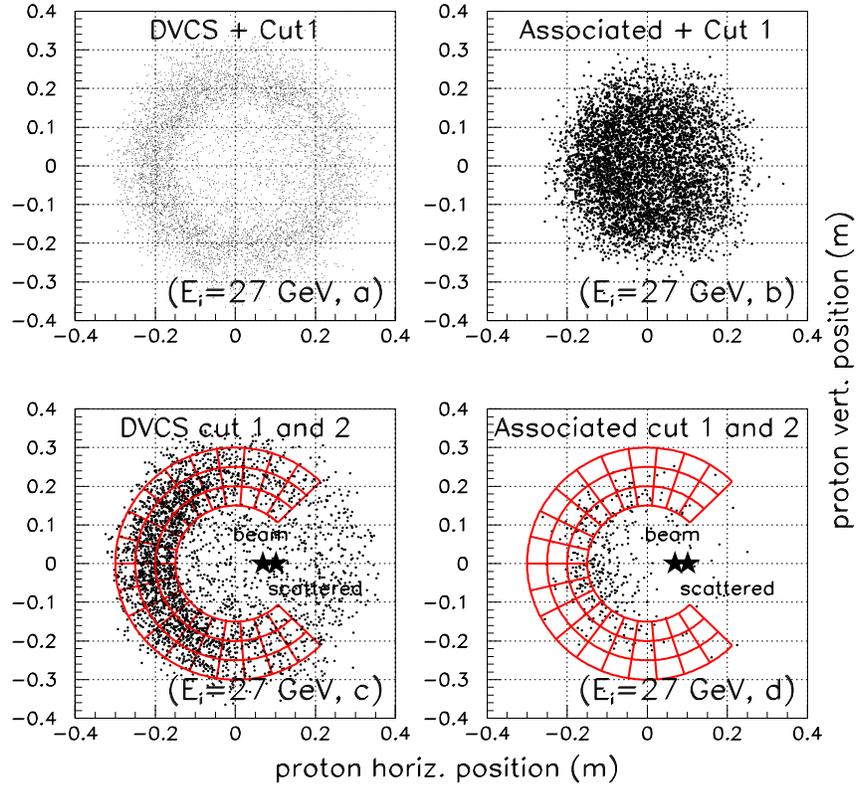,width=12 cm,height=12 cm}}
\caption{Localization of the proton in a plane 
perpendicular to the mean virtual photon direction,
 30 cm from the LH2 Target. Kinematic
 is at Q$^2$=5 GeV$^2$, $s=11$ GeV$^2$,  27 GeV energy incident. The electron spectrometer  is a new spectrometer
 at 5.7 degree with
 a solid angle of 2 msr. The inner photon calorimeter is located at -12.9 degrees and this solid angle is 40 msr.
 Note that the density of the event is just given by the phase space. }
\label{calo_position_27}
\end{figure}

Figs.~\ref{calo_position_8},\ref{calo_position_27}, show the intercept of 
the
recoil protons with the plane defined by the array of scintillators.
Plots a) and c) show DVCS events,  plots b) and d) show the associated 
production events.  Plots a) and b) are for all of the the corresponding
events, Plots c) and d) are after the cut on missing mass from 
Figs.~\ref{cut_ei8},\ref{cut27} is applied.
The protons of the DVCS and the
 associated events do not have the same distribution. The phase space (4 body)
  of the
 associated production events is  evenly spread on the acceptance whereas the 
 DVCS events are concentrated at a ring on the edge of the phase space. This is even more marked if we require a cut on the missing mass obtained with the electron and the photon.

\begin{figure}
\centerline{\epsfig{file=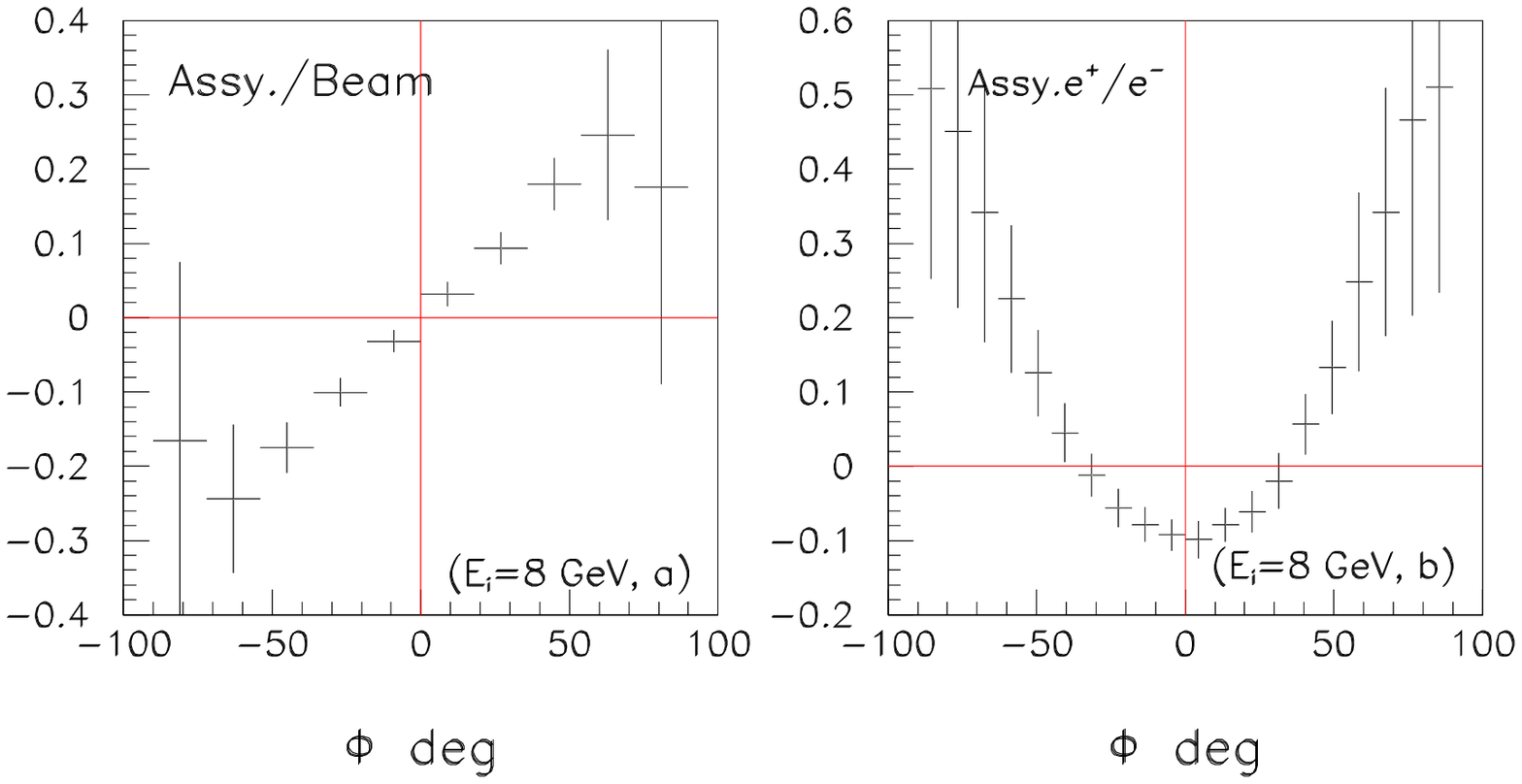,width=2truein,width=12 cm,height=6 cm}}
\vskip-0.5cm
\caption{Expected asymmetry (a) with the polarization of the beam (b) with the beam
 charge,  in 400h Hour of beam time at a Luminosity
 of 10$^{37}$ cm$^{-2}$s$^{-1}$. Kinematic is at Q$^2$=3GeV$^2$, S=8 GeV$^2$,  8 GeV 
energy incident. The electron spectrometer HMS is at 19.9 degrees with a solid angle
 of 8 msr. The inner photon calorimeter is located at -11.9 degrees and this solid
 angle is 40 msr.  The definition of $\phi$ is the same as in
 Figs.~\ref{Asymmetry ee},\ref{Asymmetry phi}}
\label{asymetri_ei8}
\end{figure}

\begin{figure}
\centerline{\epsfig{file=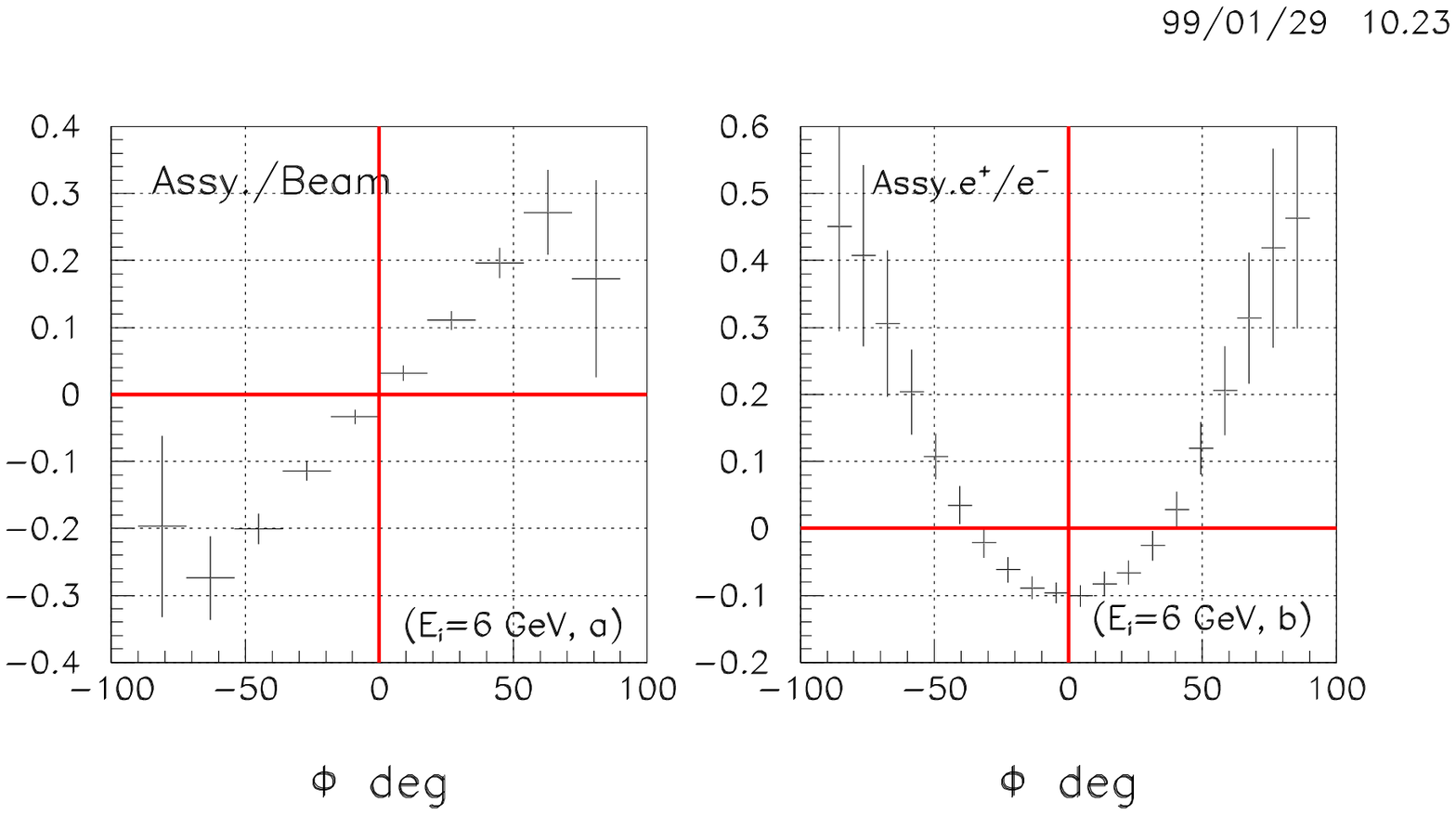,width=2truein,width=12 cm,height=6 cm}}
\vskip-0.5cm
\caption{Expected asymmetry (a) with the polarization of the beam (b)
 with the beam
 charge,  in 400h Hour of beam time at a Luminosity
 of 10$^{37}$ cm$^{-2}$s$^{-1}$. Kinematic is at Q$^2$=2 GeV$^2$, S=6 GeV$^2$,  6 GeV
 energy incident. The electron spectrometer HMS is at 22.4 degrees with a solid angle
 of 8 msr. The inner photon calorimeter is located at -11.9 degrees and its solid
 angle is 40 msr }
\label{asymetri_ei6q2}
\end{figure}

\begin{figure}
\centerline{\epsfig{file=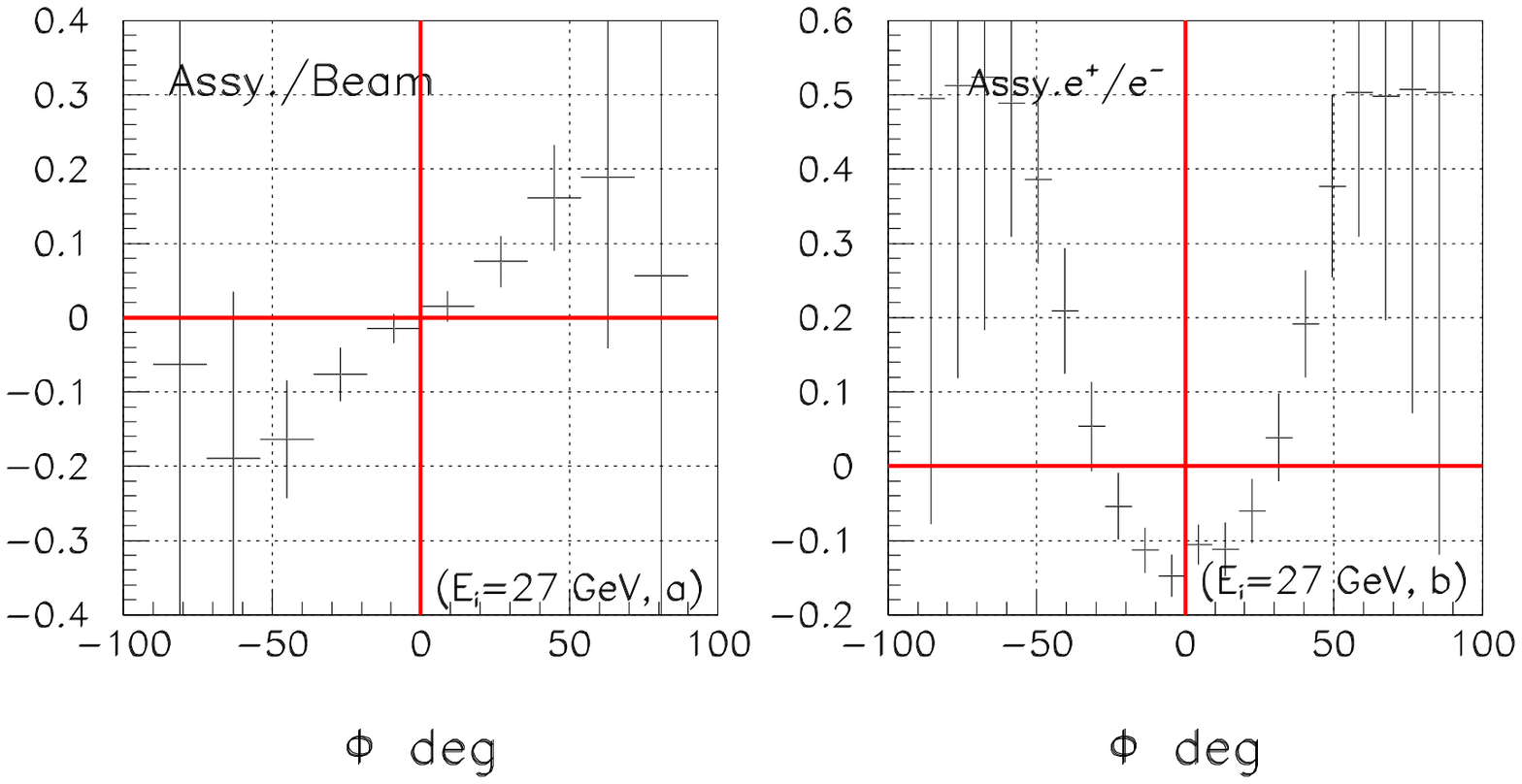,width=2truein,width=12 cm,height=6 cm}}
\vskip-0.5cm
\caption{Expected asymmetry (a) with the polarization of the beam (b) with the beam
 charge,  in 400h Hour of beam time at a Luminosity
 of 10$^{37}$ cm$^{-2}$s$^{-1}$.  Kinematic is at Q$^2$=5 GeV$^2$, S=11 GeV$^2$,  27 GeV
 energy incident. The electron spectrometer  is new spectrometer at 5.7 degrees with a solid angle
 of 2 msr. The inner photon calorimeter is located at -12.9 degrees and its solid
 angle is 40 msr }
\label{asymetri_ei27q5}
\end{figure}

Fig.~\ref{asymetri_ei8},\ref{asymetri_ei6q2},\ref{asymetri_ei27q5}{}
 give for three kinematics the expected yield for measurements
of the beam-charge and beam-helicity asymmetries.
This simulation  was done using the code ``BITCH'' \cite{BITCH},
which  
 takes into account the resolution of the spectrometer, 
the multiple scattering 
in the target, and the DVCS cross section model  of 
Ref.~\cite{PAM Guichon and M. Vanderhagen}.
 It must be noted that in the two lowest energy settings (6 and 8 GeV) the electron spectrometer
 corresponds to the CEBAF HRS spectrometer. At 16 and 27 GeV the angular
 acceptance is much smaller. The calorimeter is a lead glass 
calorimeter with a preshower compensation. 
 No attempt was made to optimize the size of this calorimeter for counting rate (at 6 and 8 GeV the counting 
rate can be increased by using a bigger calorimeter surface.)
%
These asymmetries 
contain the physics of the off forward parton
distributions.  The  $Q^2$ evolution (at fixed $x_{\rm Bj}$) of 
these asymmetries is a direct test of the 
theoretical framework of DVCS, independent of any model
of the OFPD \cite{M.Diehl Ralston and B. Pire}.

\section{Experimental Details for Large $P_T$ physics}

 In  Table~\ref{Table:PT}, we give the counting rate by day for several incident energies and kinematics.
 Since  the large $P_T$ domain can be reached by two symmetrical kinematics around the virtual photon it
 is possible to place the proton spectrometer in the same side  
(relative to the electron beam)
(positive angles)  the electron spectrometer.  The photon calorimeter is then placed in the opposite 
hemisphere  (negative angle). We can chose a center of mass angle not too far from 90$^0$  which maximize the  
$P_T$.  
 If we want to increase the solid angle, (linked to the proton jacobian  
 ${d\Omega^{cm}}/{d\Omega_p}$) we can put the calorimeter in the backward direction. This will also decrease the electro magnetic background allowing us to work at higher luminosities.

\par

 We have assumed  a low angular acceptance of the electron spectrometer (2.5 msr)   compatible with
 the small angles. Since the proton spectrometer is at larger angle we have 
 taken the HRS CEBAF spectrometers solid angle acceptance.
 
\par 
The purpose of the photon calorimeter is only to add a third arm coincidence to reject accidentals.
 Therefore, it does not need high energy resolution. It can be done with
 a sandwich of lead and plastic scintillators. A lead sheet placed in front to protect it from low energy X-ray and photons. We will then just set a threshold on this calorimeter response to get rid of the low energy noise.

 The photon calorimeter  angular range is large, since it has to match the proton acceptance.
 $d\Omega_{\gamma}=d\Omega_p\times(\frac{d\Omega^{cm}}{d\Omega_p})(\frac{d\Omega_\gamma}{d\Omega^{cm}})$.
 \par
 For the large  $P_T$ reaction the associated BH amplitude  is small, and will be not enhanced by ABH. This means it will be smaller than in the DVCS case so it should not be a problem.
 On the other hand the pion electro production is becoming relevant in this 
 kinematic. We will use  missing mass cuts to select our events. 
 The missing mass will be constructed with the scattered electron and the 
 recoil proton.
If this is not enough, then  we can use the photon calorimeter granularity 
for coplanarity cuts. The 
 acoplanar events  will be used to obtain the $\pi^0$ cross section. 
 We will then subtract their contribution to the coplanar events to get a clean signal.
    
 \par The counting rate are given by day. It must be noted that in this example we tried 
to reach the highest s and the biggest Q$^2$.  However the cross section and the counting rate
 decrease as $s^{-6}$.   Going from $s=15$ GeV$^2$ to $s=12$ GeV$^2$ increases the counting
 rate by a factor 3.8. Decreasing Q$^2$ increases the counting rate also. 

\begin{center}
\begin{table}[ht]
\begin{tabular}{||c|c|c|c|c|c|c|c|c||} \hline \hline
$\theta^{\gamma\gamma}_{cm}$&$P_{T}$&
 $\theta^{lab}_\gamma$& $\theta^{lab}_p$&$P_p$&
\rule[-2pt]{0pt}{14pt} ${d^5\sigma \over dP_e\, d\Omega^{cm} }$& 
$\frac{d\Omega^{cm}}{d\Omega_{\gamma}^{\rm lab} }$&$\frac{d\Omega^{cm}}{d\Omega_p^{\rm lab}}$&N \\
{deg}&{GeV}&
{deg}&{deg}&{GeV}&
\rule[-6pt]{0pt}{16pt}${\rm pb\over GeV\, sr^{2}}$&&&{day$^{-1}$} \\ \hline \hline
\multicolumn{9}{|l|}{\rule[-2pt]{0pt}{14pt}$E_i = 8$ GeV, $s = 8$ GeV$^2$,
   $Q^2 = 0.5$ GeV$^2$, $\theta_e = 7.22^\circ$,  $p_e = 3.9$ GeV}\\ \hline
 90&1.25&$+27.6$&$-35.7$&2.22&.406& 3.12& 5.1&120 \\  
120&1.08&$+49.6$&$-24.7$&3.57&.406& 1.07& 8.5&200 \\ \hline \hline
\multicolumn{9}{|l|}{\rule[-2pt]{0pt}{14pt}$E_i = 12$ GeV, $s = 10$ GeV$^2$,
      $Q^2 = 1.0$ GeV$^2$,  $\theta_e = 6.43^\circ$,  $p_e = 6.6$ GeV}\\ \hline
 90&1.44&$+22.2$&$-33.5$&3.31&.139& 4.0& 6.1& 82 \\ 
120&1.24&$+42.0$&$-23.5$&4.60&.139& 1.28&10.6&149 \\ \hline\hline
\multicolumn{9}{|l|}{\rule[-2pt]{0pt}{14pt}$E_i = 27$ GeV, $s = 11$ GeV$^2$,
   $Q^2 = 5.0$ GeV$^2$,  $\theta_e = 5.7^\circ$,  $p_e = 18.9$ GeV}\\ \hline
 60&1.32&$-25.5$&$+15.4$&2.78&.085&15.7& 4.9&149 \\ 
 70&1.43&$-28.2$&$+11.9$&3.41&.085&12.8& 6.6&160 \\ \hline \hline
\multicolumn{9}{|l|}{\rule[-2pt]{0pt}{14pt}$E_i = 27$ GeV, 
$s = 15$ GeV$^2$, $Q^2 = 2.0$ GeV$^2$, 
$\theta_e = 3.6^\circ$,  $p_e = 18.4$ GeV}\\ \hline
120&1.58&$+32.7$&$-20.7$&7.03&.041& 1.78&15.4&175\\ 
 90&1.82&$-31.7$&$+13.9$&4.96&.041& 6.05& 8.2& 96\\ \hline \hline
\end{tabular}
\caption{Large P$_T$: Counts obtained in 1 day, with
  a beam of 80. $\mu$A,  10 cm of liquid hydrogen target,
an electron spectrometer of 2.5 mr and a momentum acceptance of $\pm $2.5\% 
and an acceptance of 7 mr for the proton spectrometer. 
The detection of photon is assumed to match the acceptance of the proton 
spectrometer.
}
\label{Table:PT}  
\end{table}
\end{center}

\section{Conclusion} 
We have shown that experimentally from 6 to 27 GeV it is 
possible to
 access the VCS, the DVCS and the VCS at Large $P_T$.
 For the DVCS  up to $s \le 15 $GeV$^2$ and $Q^2 \le 7$ GeV$^2$ 
 (for $x_{\rm Bj} \approx 0.3$).  
We have also shown that for the large $P_T$ VCS ($P_T=1.9$ GeV) 
we can reach  $s= 15$ GeV$^2$ 
and $Q^2$=2 GeV$^2$.
 We have also shown that the apparatus (detector) is feasible.  We already did 
 experiments  
 at this level of accuracy (CEBAF), momentum (SLAC) and background  (RCS studies
 at CEBAF).
 It is also evident that a positron beam would be a big advantage for DVCS.  We believe
 that it is technically possible to have such a beam at CEBAF. 
 It is almost a requirement for the new ELFE machine to be able to deliver 
 both electrons and positrons beams since so much physics can benefit from it.

\section{Acknowledgments}
We thank P.A.M. Guichon, M. Vanderhaeghen, A.V. Radyushkin, and M. Diehl
for many fruitful discussions, and for their interest in developing this
physics.  We want to point out especially P.A.M. Guichon who gave us the code
we used to evaluate the DVCS and BH processes.  P.Y. B. and Y. R. have done
this work for the European network (HaPHEEP). The work of C.E.H.-W. was
supported by the U.S. Dept of Energy and N.S.F.

\end{document}